\documentclass[fp,twocolumn]{jpsj3}
\usepackage{txfonts}

\usepackage{graphics,color,url}

\title{Navigating the Trends in the Proximity to
Magnetic Quantum Critical Point and Superconductivity for Ce-Based Heavy-Fermion Compounds}

\author{Munehisa Matsumoto$^{1,2}$\thanks{
    Present address:
Institute of Materials Structure Science,
High Energy Accelerator Research Organization (KEK),
Oho 1-1, Tsukuba, Ibaraki 305-0801, Japan
}}

\inst{
$^{1}$Institute for Solid State Physics (ISSP), University of Tokyo,
Kashiwa-no-ha 5-1-5, Kashiwa 277-8581, JAPAN\\
$^{2}$Elements Strategy Initiative Center for Magnetic Materials (ESICMM),
National Institute for Materials Science (NIMS),
Sengen 1-2-1, Tsukuba 305-0047, JAPAN
}

\date{\today}

\abst{
Pressure-induced magnetic quantum phase transition of CeRhIn$_5$
is described from first principles based on a realistic Kondo lattice model.
Simulations of CeRhIn$_5$ above ambient pressure up to $P=5$ [GPa]
show that it is driven from an antiferromagnetic phase
into the paramagnetic phase crossing the possible QCP
around $P=2$ [GPa], being consistent with the experimental trends.
Our calculations suggest a possibility for a coexistence phase around $P\stackrel{<}{\sim} 3$ [GPa].
On the basis of the data from our realistic simulations,
a comprehensive materials navigator is compiled including pressure dependent CeRhIn$_5$
to possibly provide a guideline for further discovery of Ce-based heavy fermion superconductors.
}

\begin{document}
\maketitle

\section{Introduction}

\begin{figure}
\scalebox{0.8}{\includegraphics{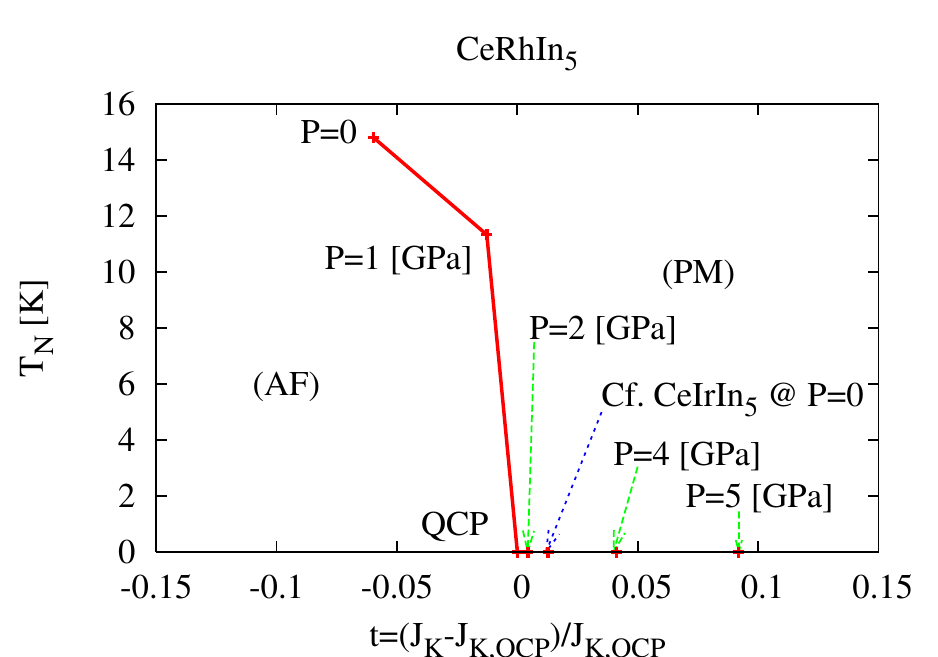}}
\caption{\label{doniach} (Color online)
  Rescaled Doniach phase diagram to show CeRhIn$_5$ under several
pressures residing in the neighborhood of the magnetic QCP. The data for CeRhIn$_5$
and CeIrIn$_5$ at ambient pressure~\cite{mm_2010} are also shown for comparison.}
\end{figure}

The Ce-115 heavy-fermion materials have been
attracting much attention since their discovery at the beginning of the 21$^{\rm st}$ century
 due to its relatively high superconducting transition temperature
$T_{\rm c}$ among Cerium-based heavy fermion materials. CeCoIn$_5$ at ambient pressure
has $T_{\rm c}=2.3$~[K]~\cite{petrovic_2001} which rises slightly under pressure,
and CeRhIn$_5$ under pressure $P\sim 2$~[GPa] shows
$T_{\rm c}=2.1$~[K] ~\cite{hegger_2000}. These are much higher than $ T_{\rm c}$'s
of other typical heavy fermion superconductor such as CeCu$_2$Si$_2$ of which $T_{\rm c}=0.7$~[K]~\cite{steglich_1979}.
A Pu variant of 115-material family, PuCoGa$_5$, has $T_{\rm c}$
as high as $18.5$~[K]~\cite{sarrao_2002}.
This route of rising $T_{\rm c}$'s looks promising toward the possible peak.
Understanding Ce-115's has been one of the central issues up to today~\cite{shishido_2018}.
Being in the proximity to a magnetic quantum critical point (QCP)~\cite{sachdev_1999},
where magnetic long-range order collapses and associated fluctuations are diverging,
is thought to control many interesting phenomena including superconductivity. Addressing QCP
is an important issue to understand the relatively
high-$T_{\rm c}$ in Ce-based heavy fermion compounds.
In this work we present {\it ab initio} description of pressure-induced QCP in CeRhIn$_5$
and let the calculated results overlap
with an overall trends of related compounds
to compile a materials navigator for Ce compounds around a magnetic QCP. The main results are shown in
Fig.~\ref{doniach} for the pressure-induced
QCP in CeRhIn$_5$ as found around $P\sim 2$ [GPa] in good agreement
with experiments~\cite{shishido_2005}.
The materials navigator is shown in Fig.~\ref{fig::navi} below.

In the next section we outline our methods.
In Sec.~\ref{sec::results}, some aspects of the results presented in Fig.~\ref{doniach}
are highlighted and the comprehensive materials navigator is shown. This is followed
by brief discussions about what has been included in the descriptors of the navigator
as the dominating factors of the overall trend and what has been dropped.
Final section is devoted for conclusions and outlook.

\section{Methods}
\label{sec::methods}

\begin{figure}
\scalebox{0.8}{\includegraphics{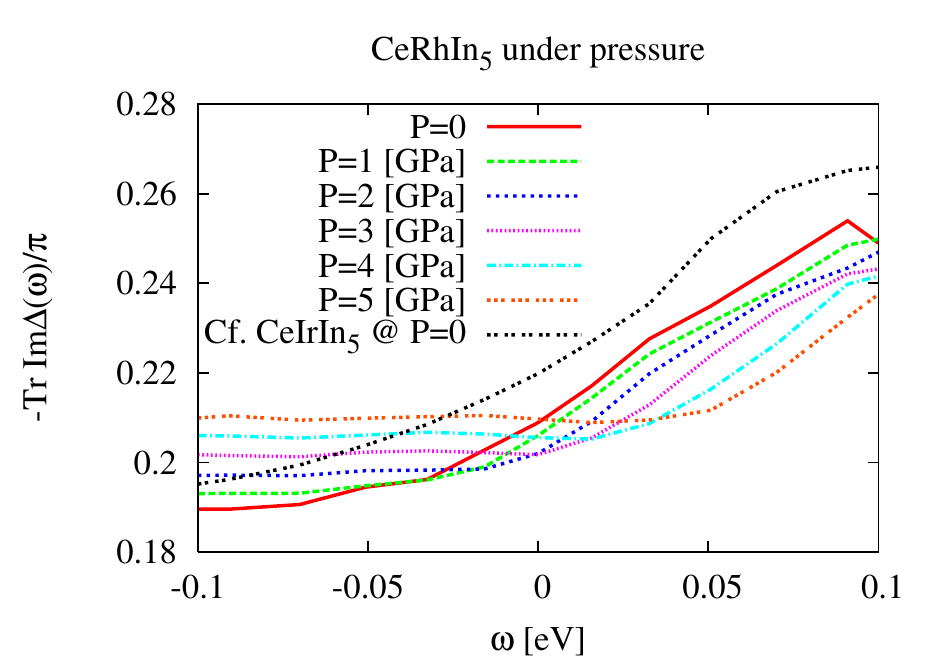}}
\caption{\label{ImDelta} (Color online)
  LDA+DMFT (Hubbard-I) result for
the hybridization function of CeRhIn$_5$ under pressures.
The data for CeRhIn$_5$ and CeIrIn$_5$ at ambient pressure~\cite{mm_2010}
is overlapped for comparison. The data is shown focusing
around the Fermi level.}
\end{figure}
We locate the magnetic QCP by looking at the Doniach phase diagram~\cite{doniach_1977}
of a realistic Kondo lattice model (KLM) defined for a given material from first-principles.
Following Ref.~\citen{mm_2009}, the hybridization function between the localized
Cerium $4f$-orbital and conduction electron band composed of $s$, $p$, and $d$-orbitals
defines the realistic Hamiltonian for the given material. Then the magnetic phase transitions can
be tracked by sweeping the Kondo coupling $J_{\rm K}$, and at some point in $J_{\rm K}$
the magnetic transition temperature vanishes, reflecting the competition between the magnetic ordering energy scale
and the Kondo screening energy scale~\cite{doniach_1977}. We identify the boundary between the antiferromagnetic phase
and the paramagnetic phase to be the magnetic QCP, with a certain reserve regarding a possibly subtle
characterization of QCP. The local density approximation (LDA) combined with dynamical mean field theory (DMFT)
(Hubbard-I)~\cite{gabi_2006} results for the hybridization function, $-\Im\Delta(\omega)/\pi$,
for CeRhIn$_5$ under pressures,
together with CeRhIn$_5$ and CeIrIn$_5$ at ambient pressure~\cite{mm_2010},
are shown in Fig.~\ref{ImDelta}. 
For the material-specific calculation of the hybridization function,
we have used full-potential linear muffin-tin orbital method~\cite{sergey}
as implemented in LmtART~\cite{weblink}.
The hybridization at the Fermi level is most important~\cite{burdin_2000}
for the Kondo screening and it is seen that CeIrIn$_5$ has stronger hybridization
than CeRhIn$_5$, which would favor the Kondo screening for CeIrIn$_5$. We also
see that the pressure dependence of the value of $-\Im\Delta(0)/\pi$ for CeRhIn$_5$
is non-monotonic and the energy-dependence of the hybridization function would also
be relevant in determining the fate of the magnetism in CeRhIn$_5$ under pressure.

Thus defined KLM is solved up to the approximation of DMFT~\cite{georges_1996}
using the continuous-time quantum Monte Carlo solver~\cite{rubtsov_2005,werner_2006,haule_2007}
for the Kondo impurity problem~\cite{otsuki_2007}. In this impurity problem
we incorporate the realistic crystal-field and spin-orbit level splittings on the
local $4f$ orbital of Cerium. For Ce-115's, it is known that the tetragonal crystal structure splits
the $j=5/2$ multiplets into three doublets, separated by $\Delta_{1}=5.9$~[meV] between the lowest
${\Gamma}_{7}=$ and the second-lowest $\Gamma_{7}$, and $\Delta_{2}=28.5$~[meV]
between the lowest $\Gamma_{7}$ and the highest $\Gamma_{6}$ for CeRhIn$_5$~\cite{takeuchi_2001,latest}.
For CeIrIn$_5$ the splittings is $\Delta_{1}=5.26$~[meV] and $\Delta_{2}=25.875$~[meV]~\cite{takeuchi_2001,latest}.
We set the spin-orbit splitting between $j=5/2$ and $j=7/2$ multiplets to be $\Delta_{\rm spin-orbit}=0.3$~[eV].

The lattice parameters are taken from experimental values~\cite{rmp_2009}
and the bulk modulus for CeRhIn$_5$ under pressure have been taken from Ref.~\citen{normile_2005}.
Here we look at the effects of hydrostatic pressure and take into account only the volume
compression effects, keeping the ratio of the lattice constants, $c/a$, to be constant.
The effects of the variation of $c/a$ would enter by orders of magnitude smaller.
In LDA, we also vary the local $4f$-level, $\epsilon_{f}$, to better
track the pressure evolution of energy scales, instead of fixing it to $\epsilon_{f}=-2.5$ [eV]
as we did in our previous work~\cite{mm_2009}.
Fixing $\epsilon_{f}$ works fine as far as the given
material stays in a very good Kondo limit.
However to tract the pressure evolution
of energy scales in Ce-115's, we find that adjusting $\epsilon_{f}$ seems to be essential~\cite{pourovskii_2008}.
The values of the volume compression ratio and the output of LDA for $\epsilon_{f}$
are summarized in Table~\ref{numbers}.
\begin{table}
\caption{\label{numbers}
The parameters for CeRhIn$_5$ under each pressure and the result
for the Kondo temperature of the single-impurity problem.
Shown are the volume compression ratio $V/V_{0}$
following Ref.~\citen{normile_2005},
the local $f$-level position $\epsilon_{f}$ obtained with
LDA + DMFT (Hubbard I), the value of the hybridization function on the Fermi level
summed over the $4f$-orbitals, $-{\rm Tr}\Im\Delta(0)/\pi$, and the result for
the single-ion Kondo temperature. The input data of $\epsilon_{f}$
and $-{\rm Tr}\Im\Delta(0)/\pi$ for the ambient-pressure materials~\cite{mm_2010}
are also shown for comparison.
}
\begin{tabular}{rcccc}\hline
$P$ [GPa] & $V/V_{0}$ & $\epsilon_{f}$ [eV] & $-{\rm Tr}\Im\Delta(0)/\pi$ [eV] & $T_{\rm K}^{0}$ [K]\\ \hline
CeRhIn$_5$ at $P=1$ & 0.98863 & -1.83 & 0.206 & 5.4 \\ \hline
$2$ & 0.97789 & -1.77 & 0.202 & 5.9 \\ \hline
$3$ & 0.96771 & -1.71 & 0.202 & 6.6 \\ \hline
$4$ & 0.95806 & -1.65 & 0.206 & 7.4 \\ \hline
$5$ & 0.94888 & -1.59 & 0.210 & 8.1 \\ \hline\hline
Cf. CeRhIn$_5$ at $P=0$ & 1 & -1.90 & 0.209 & 4.9 \\ \hline
Cf. CeIrIn$_5$ at $P=0$ & 1 & -1.95 & 0.220 & 6.0 \\ \hline
\end{tabular}
\end{table}

\begin{figure}
\scalebox{0.8}{\includegraphics{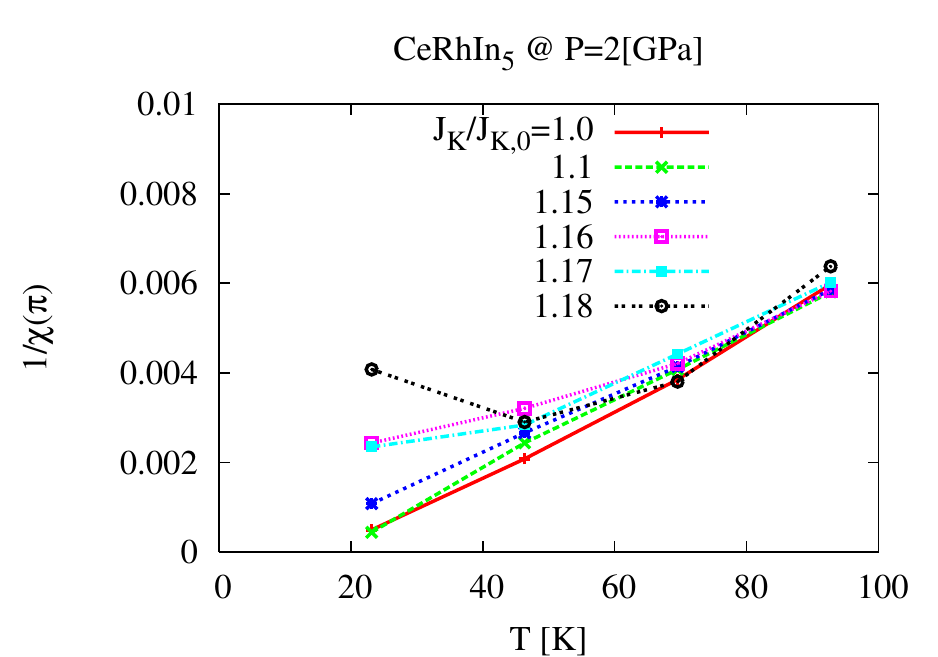}}
\caption{\label{ssus} (Color online) Temperature dependence of the inverse of
the staggered magnetic susceptibility $\chi(\pi)$ of realistic Kondo lattice models
for CeRhIn$_5$ at $P=2$~[GPa]. The Kondo coupling $J_{\rm K}$ is swept from $J_{\rm K,0}$
at which the effective Hund coupling $J_{\rm Hund}=0$ to where the N\'{e}el temperature vanishes
or $\chi(\pi)$ saturates, and the realistic data point is picked up at $J_{\rm Hund}=1$~[eV].
For the present case it is $1.16 J_{{\rm K},0}$ and the data here shows this material
is slightly off the verge of collapsed magnetic phase.
}
\end{figure}

\begin{figure}\scalebox{0.8}{\includegraphics{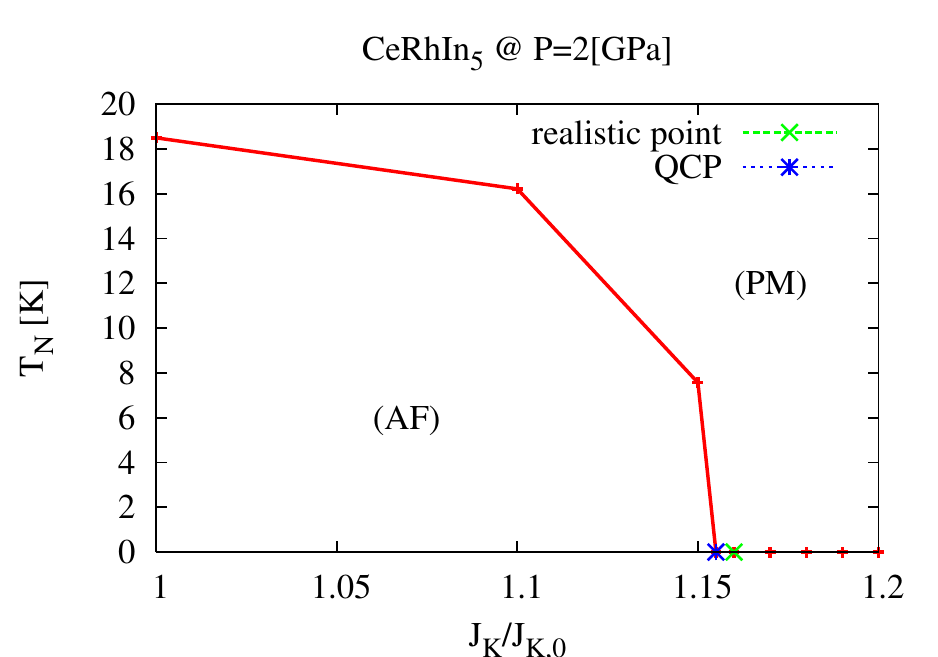}}
\caption{\label{doniach_P_2p0_GPa} (Color online) Restored Doniach phase diagram for CeRhIn$_5$
  at $P=2$~[GPa] from the data in Fig.~\ref{ssus}.
    We identify a quantum critical point (QCP) between the antiferromagnetic
phase (AF) and the paramagnetic phase (PM).}
\end{figure}

The input obtained with LDA+DMFT (Hubbard I)
to our Kondo problem is shown in Fig.~\ref{ImDelta}. Here we need an extra care for
the definition of Kondo coupling $J_{\rm K}$ because the impurity level can also shift as a pressure
effect, which comes into the definition of $J_{\rm K}$
via a realistic variant of the Schrieffer-Wolff transformation~\cite{schrieffer_1966,mm_2009}
\begin{equation}
J_{\rm K}=\int_{-\infty}^{D}d\,\omega \frac{-{\rm Tr}\Im\Delta(\omega)}{\pi N_{\rm F}}
\left
[\frac{1}{|\epsilon_{f}|}+\frac{1}{(U+\epsilon_{f}-J_{\rm Hund})}
\right],
\label{sw}
\end{equation}
where $D$ is the band cutoff set to be equal to the Coulomb repulsion $U=5$~[eV],
$N_{\rm F}=14$ is the number of orbitals and $J_{\rm Hund}$ is the effective Hund
coupling in the $f^2$ multiplet to which the second term of Eq.~(\ref{sw})
describes the virtual
excitation from the $f^1$ ground state. Practically we sweep $J_{\rm Hund}$, and
thus $J_{\rm K}$, to locate the QCP on a Doniach phase diagram and also to pick
up the realistic data point at $J_{\rm Hund}=1$~[eV]. In this way we can see
where in the neighborhood of QCP our target material resides on the Doniach phase
diagram. The temperature dependence of the staggered magnetic susceptibility $\chi(\pi)$
is observed for each $J_{\rm K}$ and we extrapolate it linearly to the low temperature
region to see if there is a finite N\'{e}el temperature.
For the calculation of two-particle Green's function
we follow the formalism developed in Ref.~\citen{otsuki_2009_formalism}
and we use a random-dispersion approximation to decouple it into single-particle
Green's functions~\cite{gebhard_1997}, which would enhance the transition temperature
together with DMFT.
The data is shown in Fig.~\ref{ssus} for the case of CeRhIn$_5$ at $P=2$~[GPa]
by which we identify
that the N\'{e}el temperature vanishes in the
parameter range $1.15J_{{\rm K},0}<J_{\rm K}<1.16J_{{\rm K},0}$, where $J_{{\rm K},0}$
is the Kondo coupling at $J_{\rm Hund}=0$. Actually the realistic data point
is obtained by plugging in $J_{\rm Hund}=1$~[eV] and $\epsilon_{f}=-1.77$~[eV]
(as can be found in Table~\ref{numbers}) to Eq.~\ref{sw} to be $J_{\rm K}=1.16J_{\rm K,0}$.
So the data in Fig.~\ref{ssus} shows that CeRhIn$_5$ at $P=2$~[GPa] is almost right
on the magnetic QCP. The restored Doniach phase diagram for this case is shown in
Fig.~\ref{doniach_P_2p0_GPa}. The same analysis procedure is applied to CeRhIn$_5$
under all investigated pressures here to lead to the main result in Fig.~\ref{doniach}.
There we have rescaled the horizontal axis of the Doniach phase diagram to look at
the distance to QCP independently of the material (or pressure)-specific energy scale.

\section{Results}
\label{sec::results}

\subsection{Pressurized CeRhIn$_5$}

\subsubsection{Quantum Phase Transitions}

\begin{figure}
\scalebox{0.8}{\includegraphics{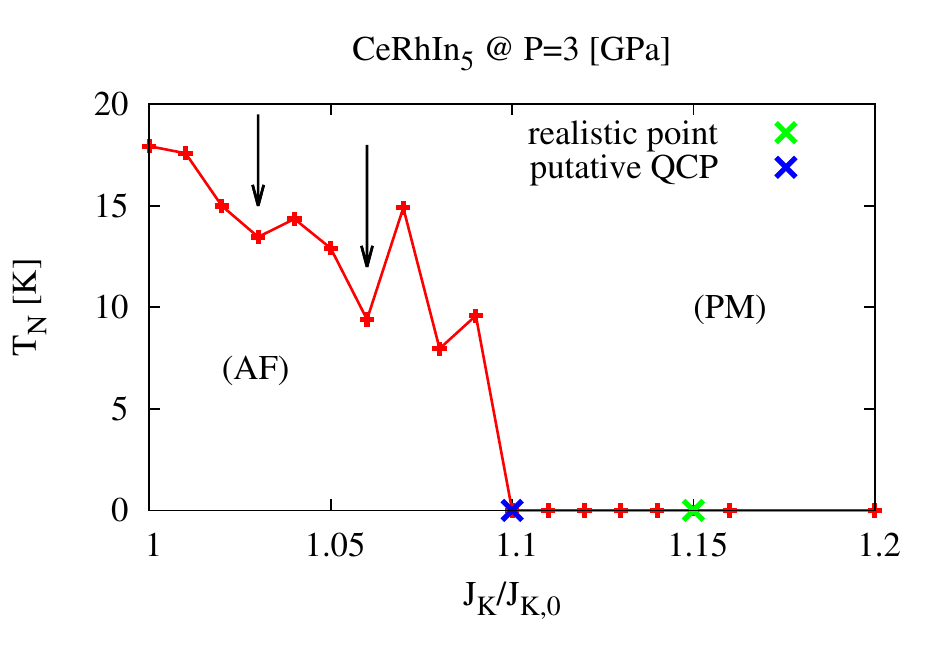}}
\caption{\label{doniach_P_3p0_GPa}(Color online)
  Realistic Doniach phase diagram
  obtained for CeRhIn$_5$ at $P=3$ [GPa].
Between the antiferromagnetic
phase (AF) and the paramagnetic phase (PM), we encounter
a parameter region where a competing trend between an approach to
quantum criticality and an enhancement of the antiferromagnetic order
is numerically observed. The putative trend to quantum criticality is
indicated by arrows in the figure on the region
where a suppression of calculated N\'{e}el temperatures is seen.
}
\end{figure}

The result for $P=3$~[GPa], which is expected to be
slightly into the paramagnetic
phase crossing the magnetic QCP as it should be somewhere between $P=2$~[GPa]
and $P=4$~[GPa] in Fig.~\ref{doniach},
is worth inspecting closely even though the restored Doniach phase diagrams
might look similar in many cases. The restored Doniach phase diagram for this case
is shown in Fig.~\ref{doniach_P_3p0_GPa}.
We find a certain parameter range on the $J_{\rm K}$ axis
near a putative QCP where non-monotonic trend between the antiferromagnetic phase
and the paramagnetic phase is numerically observed.
Here we note that in the analysis procedure described in the previous section,
we just identified QCP by a parameter segment where the
finite N\'{e}el temperature seems to have vanished.
In principle, there always remains a possibility that
the calculated N\'{e}el temperature jumps at some point or oscillates 
in some smaller parameter segments than is actually inspected.
Among the pressure data points
investigated here, only for the case of $P=3$~[GPa] the coexistence region seems to be
wide enough to be detected within the current numerical resolution.
This could mean that there is a coexistence phase or a first-order
transition point around what has been called QCP in the pressure range $P\stackrel{<}{\sim} 3$~[GPa],
considering that the artificial factor on Kondo temperature can be qualitatively
regarded as an effective pressure around a given material-specific parameter.
We leave the exact characterization of what we also call QCP here for future investigations:
our main point here is that a realistic modeling of CeRhIn$_5$ under pressure
gives the trend of magnetism around QCP
in a semi-quantitative agreement with experimental observations.
For the moment
we would be satisfied with that it looks like
a continuous disappearance of calculated N\'{e}el temperatures
to be practically considered as a QCP
in most cases even with a very small numerical jump.
Also it is worth noting that experimentally it was observed
under the strong magnetic field that that the Fermi surface jumps
from a small one to the large one around the magnetic QCP
at $P_{\rm c}=2.3$~[GPa]~\cite{shishido_2005}, which may point to
the true nature of what has been called QCP.
These situations motivates us for a possible microscopic origin of the two-fluid phenomenology
of heavy fermion materials~\cite{two_fluid}.

\subsubsection{Kondo Temperatures}

We now look at the pressure-evolution of the Kondo-screening energy scale.
We have extracted the single-ion Kondo temperature $T_{\rm K}^{0}$
by fitting $1/(T+T_{\rm K}^{0})$ to the high-temperature tail of the temperature dependence
of the local susceptibility of the impurity problem, or equivalently the first iteration
of DMFT,
in the temperature range
$0.001 {\rm [eV]}\leq T \leq 0.01 {\rm [eV]}$.
This procedure is not quite exact considering the crossover effect
between different degrees of the effective degeneracy
of the local $f$-orbital caused by the crystal field/spin-orbit level splitting
in the corresponding temperature range~\cite{yamada_1984, cox_1998}.
The results are summarized in the rightmost column of Table.~\ref{numbers}. The pressure evolution of the Kondo temperature
in CeRhIn$_5$ has been observed in qualitative agreement with the experimental trend
but the evolution is not as strong as exponential as was shown in Ref.~\citen{yang_2008}.
Further comparison with a refined extraction procedure of Kondo temperatures,
e.g. trying to reach the complete Pauli-paramagnetic behavior in the low-temperature limit,
would give a more reliable result at a much bigger computational cost.

\subsubsection{Quasiparticle Renormalization Factor}

\begin{figure}\scalebox{0.8}{\includegraphics{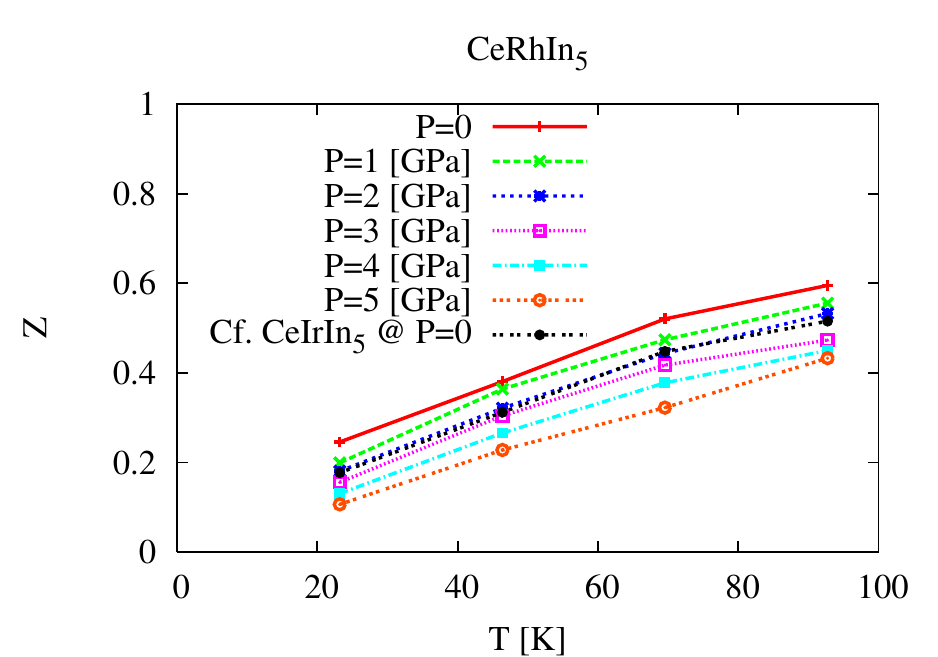}}\\
\caption{\label{CeRhIn5_Z_vs_T_and_P} (Color online) Temperature evolution of the conduction-electron
renormalization factor for CeRhIn$_5$ for various pressures. Interestingly the result at the quantum critical
point (QCP) of CeRhIn$_5$ almost collapses on that of CeIrIn$_5$ at ambient pressure, which is also close
to QCP.}
\end{figure}
Finally we look at the pressure-evolution of quasiparticle weight of the conduction electrons
within the present realistic Kondo lattice description.
From the conduction-electron self energy  $\Sigma(i\omega_{n})$,
we derive the renormalization factor with $Z=1/[1-\partial\Sigma(i\omega_{n})/\partial (i\omega_{n})]$.
Here $\omega_{n}=(2n+1)\pi T$ is the fermion Matsubara frequency.
The result for $Z$ of CeRhIn$_5$ under pressures is shown in Fig.~\ref{CeRhIn5_Z_vs_T_and_P}.
It is seen that the stronger pressure makes the conduction electrons the heavier, and interestingly
the data of CeRhIn$_5$ at $P=2$~[GPa], which is almost right on QCP, comes very close to CeIrIn$_5$,
which is also close to QCP.

Further investigations of pressure effect on Ce-115's would be interesting.
An interesting feature of CeIrIn$_5$ is that $T_{\rm c}$ increases as the pressure is imposed
while the material is supposed to be going away from its magnetic QCP under pressure~\cite{kawasaki_2005}.
From our point of view, the pressure dependence can be non-trivial
at least due to the frequency dependence of the hybridization function.
In principle the Kondo temperature can even be lowered as the pressure is increased
and the possible distance to
the magnetic QCP can be elongated or shortened depending on the pressure change
of the frequency dependence of the hybridization. We also note that when
the hybridization gets too strong under strong pressures,
the valence fluctuations would be so important that
the framework based on localized $f$-electrons may not be good anymore~\cite{holmes_2004,miyake_2018}.

\subsection{Compiled Materials Navigator}

\begin{figure}
\scalebox{0.7}{
\includegraphics{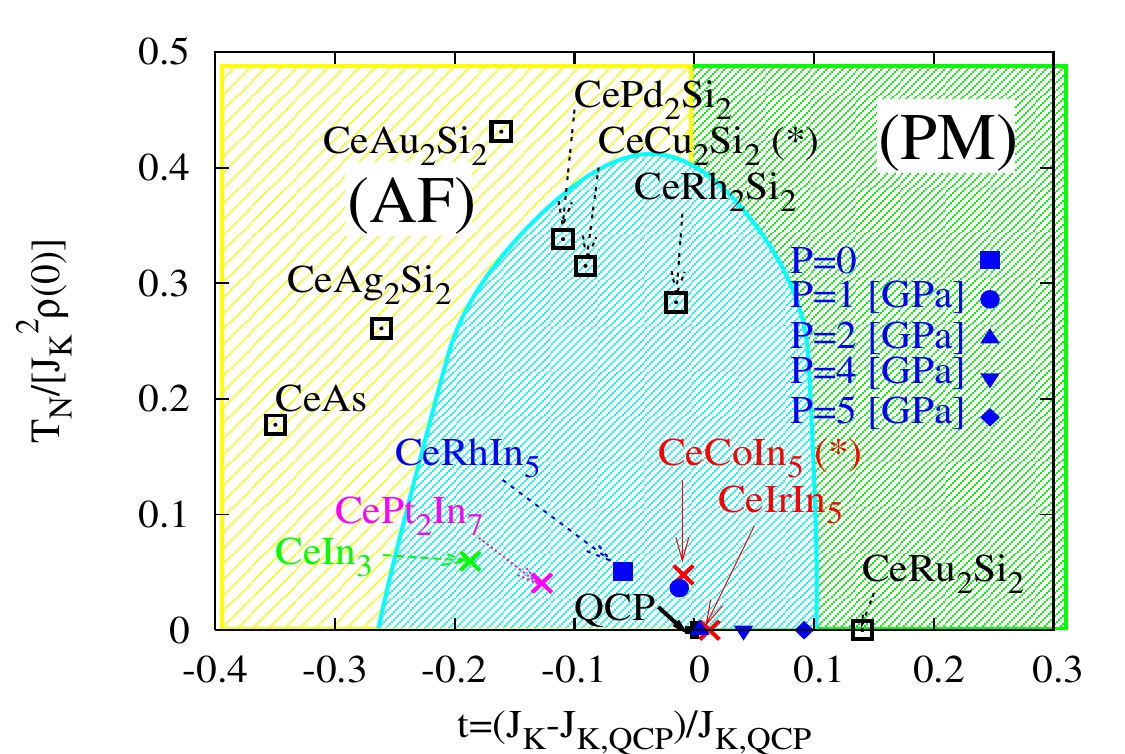}
}\\
\caption{\label{fig::navi} (Color online)
Computationally constructed
navigator for Ce-based heavy fermion compounds to
inspect the trends around the magnetic quantum critical point (QCP).
Central area around QCP harbors superconductors.
On the left-hand side antiferromagnetism (AF)
dominates and on the right-hand side paramagnetism (PM) dominates.
The asterisk (*) on the material indicates that those materials
are found in a close proximity to the QCP, where N\'{e}el temperature disappears presumably
in a continuous way but quite abruptly, and
finite the N\'{e}el temperature from the calculations may not necessarily mean that
they are really
inside the magnetic phase. The point lies in comparing
the calculated distance from the QCP among the target materials.}
\end{figure}

Rescaled data around QCP as shown in Fig.~\ref{doniach}
is further rescaled by a magnetic ordering energy scale as parametrized
by the factor for RKKY interaction and be overlapped with more analogous data
for Ce compounds as shown in Fig.~\ref{fig::navi}.
In order to eliminate possible
  material-specific energy scales of magnetic ordering, the prefactor
  of the RKKY interaction, $J_{\rm K}^2\rho(0)$, where $\rho(0)$ is the density of states
  on the Fermi level, has been used as a normalization factor for the vertical axis.
Some of the data
have been taken from
Refs.~\citen{mm_2009} and~\citen{mm_2010} for Ce-122's and Ce-115's, respectively.
Details for
Ce monopnictides including CeAs will be reported elsewhere.
It is clear that Ce-115 compounds indeed are concentrated around the magnetic QCP. Whenever a new Ce-based
compound pops up, we can calculate the realistic Doniach phase diagram and put the realistic data point
in Fig.~\ref{fig::navi}. If it turns out that
the new data point is located closer to QCP than the data points for
pre-existing superconductors, it would be likely that superconductivity is observed for the proposed material.
We note that CePd$_2$Si$_2$ is a pressure-induced superconducting material while
CeAu$_2$Si$_2$ is
  not superconducting in the range of moderate pressures up to $O(1)$GPa~\cite{jaccard_2017}.
  The possible border between plain magnetic compounds and superconductors has been empirically drawn between them
  and the possible region for a superconducting material that can be reached
  within the range of the moderate pressures is shown as the blue hatched region around the magnetic
  QCP on the origin in Fig.~\ref{fig::navi}.

\section{Discussions}
\label{sec::discussions}

Pressure-induced quantum critical point in CeRhIn$_5$ has been
reproduced from first principles in a semi-quantitative way
even within the various simplifications~\cite{mm_2009, mm_2010}
made in our computational framework. This should be due to the incorporation of 14 orbitals
and the proper inclusion of realistic level splittings together with the realistic hybridization function,
presumably putting the Kondo screening energy controlled in a close range to the experimental realities
when the characteristic energy scale of the Kondo physics scales as an exponential function
of the number of degeneracy of local impurity level. The exponential dependence on the Kondo coupling
seems to have imposed one of the biggest problems
in the past theoretical description of heavy fermion material - now that incorporation of 14 orbitals is well within reach in the modern status of realistic simulations based
on {\it ab initio} electronic structure combined with dynamical mean field theory, it is feasible to inspect the magnetism trends
among the representative compounds of Ce-based heavy fermions and superconductors as we have presented here. Within the present scope
all based on the spirit of the original Anderson model~\cite{anderson_1961} combined with the proper downfolding to the low-energy effective
model via Schrieffer-Wolff transformation~\cite{schrieffer_1966}, relative trends among magnetism of the representative heavy-fermion compounds seems to be captured as implemented on the realistic electronic structure
and we see no special need to invoke exotic mechanism to describe the location of the pressure-induced QCP in CeRhIn$_5$.

Asserting that the exponential energy scale with respect to Kondo coupling seems to have been
put under a good control, we note that we have dropped off-diagonal contribution of hybridization,
orbital-dependent hybridization, and most notably, charge fluctuations in $f$-electrons
which may become important for CeRhIn$_5$ under high pressures.
The latter has been under intense investigation~\cite{holmes_2004, miyake_2018}, and we will see in a separate
work under which parameter region the charge fluctuations contribute to what extent.
Off-diagonal contribution
from spin-orbit interaction can be difficult considering the possible negative sign problem in the quantum Monte Carlo method
used at the heart of DMFT. Even though the inclusion of the diagonal contribution in the local level splittings has yielded
the satisfactorily good results concerning the trends of N\'{e}el temperatures around pressure-induced QCP in CeRhIn$_5$,
addressing the
magnetic anisotropy takes the full incorporation of off-diagonal contribution.
This is actually important because the localized $4f$-electron in Ce brings about strong magnetic
anisotropy at low temperatures.

In the navigator shown in Fig.~\ref{fig::navi},
we have only addressed the relative trends
among antiferromagnetism, paramagnetism, and possible superconductivity
triggered around the magnetic QCP. Ferromagnetism is not covered
mostly due to the approximation
as described in Sec.~\ref{sec::methods} in
the decoupling of two-particle Green's function which basically
disables a systematic comparison between ferromagnetism and antiferromagnetism.
Incorporation of ferromagnetism and superconductivity is desirable.
It is getting feasible only recently with the development of an efficient
way to evaluate the two-particle Green's function~\cite{shinaoka_2018}.

Proper incorporation of spin-orbit interaction with the precise resolution
between ferromagnetism and antiferromagnetism is expected in the upcoming developments.

\section{Conclusions and Outlook}

We have shown that the magnetic QCP of CeRhIn$_5$ under pressure
can be quantitatively described with the realistic Kondo lattice model
based on {\it ab initio} electronic structure calculations. Near QCP
we find a numerical signature for a coexistence phase, which suggests that
1) what has been called QCP in heavy-fermion materials
could actually be a weak first-order phase transition point,
2) there is a third phase in between,
or 3) a multicritical point exists~\cite{nick}. Including the successful description
of pressure-induced quantum phase transition for CeRhIn$_5$,
a comprehensive materials navigator
has been proposed to inspect the trends between magnetism and superconductivity.
Incorporation of ferromagnetism and spin-orbit interaction poses the modern challenge.

\begin{acknowledgment}
MM's work in Institute for Solid State Physics (ISSP), Univ of Tokyo,
is supported by Toyota~Motor~Corporation.
The author gratefully acknowledges helpful discussions with H.~Shishido,
collaborations in the early stages of the present work with
J.~Otsuki, M.~J.~Han, and S.~Y.~Savrasov,
and informative interactions in DMFT workshop organized by PCoMS~\cite{PCoMS}.
This work was supported by JSPS KAKENHI Grant Number~15K13525.
Numerical calculations were executed on Numerical Materials Simulator
in National Institute for Materials Science and
ISSP Supercomputer center, Univ. of Tokyo.
\end{acknowledgment}

\end{document}